\documentclass[12pt]{article}

\usepackage[latin1]{inputenc}
\usepackage{amsfonts}
\usepackage{amsxtra}
\usepackage[margin=3cm]{geometry}
\usepackage{color}
\usepackage{hyperref}
\hypersetup{linktocpage=true}
\usepackage{graphicx}
\usepackage{mathrsfs}
\usepackage{multirow}
\usepackage{tikz}
\usepackage{amsmath,latexsym,amssymb,slashed}
\usepackage{bbm}
\usepackage{pdfpages}
\usepackage{centernot}
\usepackage{multirow}
\usepackage{varwidth}
\usepackage{slashbox}
\usepackage{amsthm}
\usepackage{xypic}

\usepackage[toc,page]{appendix}
\usepackage{bookmark}

\numberwithin{equation}{section}
\newcommand{\beq}{\begin{equation}}
\newcommand{\eeq}{\end{equation}}
\def\be {\begin{equation}}
\def\ee {\end{equation}}
\def\ba#1\ea{\begin{align}#1\end{align}}
\def\baed#1\eaed{\begin{aligned}#1\end{aligned}}
\def\bged#1\eged{\begin{gathered}#1\end{gathered}}
\def\bea{\begin{eqnarray}}
\def\eea{\end{eqnarray}}

\def\a{\alpha}

\def\e{\epsilon}
\def\ve{\varepsilon}

\def\vf{\varphi}
\def\F{\Phi}
\def\g{\gamma}
\def\G{\Gamma}

\def\k{\kappa}
\def\l{\lambda}
\def\L{\Lambda}
\def\m{\mu}
\def\n{\nu}
\def\o{\omega}
\def\O{\Omega}

\renewcommand{\t}{\theta}
\def\r{\rho}
\def\s{\sigma}

\def\z{\zeta}

\def\Tr{\text{Tr}}

\let\foo\bar
\renewcommand{\bar}[1]{ {\foo{  #1} }{} }

\newlength{\dhatheight}

\newcommand{\eq}[1]{\begin{equation}\begin{split}#1\end{split}\end{equation}}

\newcommand{\al}[1]{\begin{align}#1\end{align}}
\newcommand{\all}[1]{\begin{align*}#1\end{align*}}

\newcommand{\arxdg}[1]{\href{http://arxiv.org/abs/math/#1}{\tt math.dg/#1}}
\newcommand{\arxth}[1]{\href{http://arxiv.org/abs/hep-th/#1}{[{\tt hep-th/#1}]}}
\newcommand{\arx}[1]{[\href{http://arxiv.org/abs/#1}{\tt #1}]}

\newcommand{\acal}{\mathcal{A}}

\newcommand{\fcal}{\mathcal{F}}
\newcommand{\gcal}{\mathcal{G}}

\newcommand{\lcal}{\mathcal{L}}

\newcommand{\ncal}{\mathcal{N}}

\newcommand{\pcal}{\mathcal{P}}

\newcommand{\rbb}{\mathbbm{R}}
\newcommand{\cbb}{\mathbbm{C}}

\newcommand{\obb}{\mathbbm{1}}

\newcommand{\vt}{\vartheta}
\renewcommand{\ve}{\varepsilon}
\renewcommand{\vf}{\varphi}

\newcommand{\ea}{\bigwedge\nolimits^{\!\bullet} T^*}

\newcommand{\p}{\partial}

\newcommand{\vol}{\text{vol}}
\def\d{\text{d}}
\newcommand\nn{\nonumber}

\newcommand{\ol}[1]{\overline{#1}}

\newcommand{\ti}[1]{\tilde{#1}}

\newcommand{\cg}{\check{\gamma}}

\numberwithin{equation}{section}

\begin{document}

\baselineskip=16pt
\setlength{\parskip}{6pt}

\begin{titlepage}
\begin{minipage}{0.5\textwidth}
{\color{white} a}
\end{minipage}
\begin{minipage}{0.5\textwidth}
\flushright  
\vspace*{-.65\baselineskip}
\flushright 
\end{minipage}


\begin{center}

\vspace*{1.7cm}

{\LARGE \bf  Supersymmeric gauge theory on curved 7-branes
}

\vskip 1.6cm

\renewcommand{\thefootnote}{}

\begin{center}
 \normalsize
Dani\"el Prins
\end{center}
\vskip 0.1cm
 {Institut de physique th\'eorique, Universit\'e Paris Saclay, CNRS, CEA \\F-91191 Gif-sur-Yvette,
France}
\vskip 0.1cm
{ Dipartimento di Fisica, Universit\`{a} di Milano-Bicocca,\\
 I-20126 Milano, Italy}
\vskip 0.1cm
{ INFN, sezione di Milano-Bicocca, I-20126 Milano, Italy}
\vskip 0.2 cm
{\textsf{daniel.prins@cea.fr}}\\
\end{center}

\vskip 1.5cm
\renewcommand{\thefootnote}{\arabic{footnote}}

\begin{center} {\bf ABSTRACT } \end{center}
We construct novel $7d$ supersymmetric gauge theories which include a Chern-Simons-like term on curved spaces.
In order to so, we examine the supersymmetry constraints for E7-branes in type IIA$^*$ theory, rather than making use of an off-shell supergravity.
We find two classes of solutions to the constraints, expressed in terms of a $G_2$-structure with non-vanishing intrinsic torsion.
The supersymmetric gauge theories are then obtained by coupling flat gauge theory to such supersymmetric backgrounds.
Various examples are given, including round and squashed $S^7$ as well as $S^3 \times M_4$ with $M_4$ hyperk\"{a}hler.

\end{titlepage}

\newpage
\tableofcontents
\vspace{20pt}

\setcounter{page}{1}
\setlength{\parskip}{9pt}

\section{Introduction}
Supersymmetric field theories on Riemannian manifolds have been a topic of interest for the past decade due to the advent of localization techniques which allow for exact computations of, for example, partition functions, indices and Wilson loops, making it possible to perform checks of various highly non-trivial dualities \cite{pestun1}. The primary candidate for the Riemannian manifold underlying such a field theory is the round sphere, due to several simplifying properties: it is compact, maximally symmetric, and allows for spinors satisfying simple Killing spinor equations. As a result, a wide array of field theories on spheres has been investigated in-depth in various dimensions preserving various amounts of supersymmetry, see for example \cite{blau, pestun2, gomis, kwy, lee, kz, mz}.

More generally, field theories can preserve supersymmetry on a Riemannian manifold when the metric can be obtained as part of a supersymmetric background: a set of profiles for all fields of an appropriately chosen off-shell supergravity, such that these satisfy the supersymmetry equations of the supergravity \cite{fs}. A field theory on flat space can then be coupled to such a supersymmetric background, in such a way that the new theory is preserved under global transformations of a deformed supersymmetry algebra. The precise coupling terms can be obtained by coupling the background via a corresponding supercurrent. In this manner, classification of admissible metrics corresponds to classification of solutions of (generalized) Killing spinor equations. For example, field theories on a wide variety of manifolds have been constructed in this way for $4d$ with minimal supersymmetry \cite{js, st, ktz, dfs, lpr, df}.

Unfortunately, there are various $(d, \ncal)$ where our knowledge of off-shell supergravities and supercurrents falls short. An alternative approach to
the classification of supersymmetric backgrounds that can be coupled to field theories has been developped in \cite{triendl, mpt, mp}.
Rather than using an off-shell supergravity, it has been shown that for various cases, the Killing spinor equations that determine supersymmetric backgrounds can be reproduced by combining the type II Killing spinor equations with the kappa-symmetry constraint imposed by a supersymmetric D-brane.
This procedure leads to expressions for the off-shell supergravity fields in terms of fluxes, the dilaton and the spin connection. To some extent, this result is to be expected; on the one hand because lower-dimensional supergravities are generally obtainable as consistent truncations from type II supergravity, on the other because supersymmetric field theories on curved spaces can arise as worldvolume theories of branes wrapping manifolds with non-trivial curvature.

In this paper, we further test this string theoretic approach by examining gauge theories on curved spaces in $d=7$. There are a number of reasons for this choice of dimension. First, in $7d$, to our knowledge, a formulation for the relevant off-shell supergravity is lacking and no supercurrents are known. Secondly, since no $7d$ SCFT's exist, a brane system is a natural candidate for a UV completion to such supersymmetric gauge theories. Thirdly, $7d$ is the highest dimension for which spherical supersymmetric gauge theories with constant couplings are known \cite{blau, fhy, mz}. In addition to $S^7$, known admissible manifolds are proper $G_2$-manifolds, Sasaki-Einstein spaces and tri-Sasakians \cite{prz}. A common feature is that all of these admit spinors satisfying a (non-generalized) Killing spinor equation, similar to the sphere. However, there is no reason to expect that supersymmetric field theories do not exist on more general spaces. In general, a globally well-defined nowhere-vanishing spinor in $7d$ defines a $G_2$-structure on the manifold. The intrinsic torsion of $G_2$-structures has four components, determined by various irreducible representations of $G_2$. From this point of view, all known examples are weak $G_2$-holonomy manifolds, i.e., admitting only scalar torsion.

We will construct gauge theories on $G_2$-structure manifolds with more general torsion allowed. The theories in question are composed of a kinetic term as well as a $7d$ analogue of a Chern-Simons term, and preserve two supercharges. These theories are obtained as follows. First, we construct Killing spinor equations dictated by a supersymmetric 7-brane in a supersymmetric string background. Specifically, since the brane ought to be Euclidean, the 7-brane in question is an E7-brane in type IIA$^*$ theory \cite{hull1, hull2, hull3}. Next, we construct classes of solutions to these equations, leading to profiles for the background fields. Then we couple such background fields to the field theory and modify the supersymmetry algebra accordingly. The constructed gauge theory has an off-shell supercharge, making localization calculations feasible.

This paper is organized as follows. In section 2, we construct the Killing spinor equations. In section 3, we solve them. In section 4, we give a number of explicit examples of such solutions, including $U(1)$-squashed $S^7$ and $S^3 \times M_4$, with $M_4$ hyperk\"{a}hler. In section 5, we switch gears and describe the construction of the gauge theory. In principle, section 5 is independent from the former sections and the reader only interested in gauge theory could skip the former sections. We end with some conclusions and future directions. Appendix A contains our conventions and a number of useful identities, while explicit computations of the closure of the supersymmetry algebra and invariance of the action for the gauge theory can be found in appendix B.

\section{Killing spinor equations for E7-branes in type IIA$^*$}
In this section, we rewrite the supersymmetry constraints satisfied by Euclidean 7-branes. Our motivation comes from the procedure of \cite{fs}, where it is shown how to construct supersymmetric field theories on curved spaces by coupling a flat space supersymmetric field theory to a supersymmetric background of an off-shell supergravity. In particular, a `background' in this sense is a solution to the supersymmetry constraints of the supergravity; the supergravity equations of motions need not be satisfied.\footnote{Consider for example $\ncal =1$ field theory on $S^4$; the supergravity equations of motion are explicitly violated, as the auxiliary scalar fields play the role of the radius of the sphere and are thus non-zero \cite{st}.} In \cite{triendl, mpt, mp} it was shown how to reproduce $4d$ supergravity background constraints directly from supersymmetric branes instead. We will follow a similar procedure here for 7-branes to split the total supersymmetry constraints into two sets: the `external' and `internal' supersymmetry constraints. All components of the fluxes appear in either one or the other set, apart from the metric which determines the spin connection components in both sets. The external constraints are interpreted as the requirement that the worldvolume supersymmetry of the brane is preserved. The internal constraints are such that the brane does not break the supersymmetry of the string background. From the field theory point of view, a solution to the external constraints is sufficient to determine a curved background to which one can couple the field theory. Indeed, the external supersymmetry constraints should correspond to the supersymmetry requirements of some a priori undetermined off-shell supergravity. As we will show, we find that the external supersymmetry constraints contain the field content of a non-minimal $7d$ $\ncal =1$ supergravity coupled to a so-called `$SU(2)$-vector multiplet' \cite{strathdee}.\footnote{We adhere to the convention that $\ncal =1$ corresponds to half-maximal supersymmetry (16 supercharges) and $\ncal =2$ corresponds to maximal supersymmetry (32 supercharges).}
As far as we are aware, no off-shell formulation of such a Lorentzian supergravity is known, let alone a Euclidean variant.\footnote{
For results on $\ncal =1$ $7d$ supergravity coupled to standard vector multiplets, see \cite{tvn, chamseddine, lp, dib1}.}
Hence it is currently not possible to compare the external constraints to the supersymmetry constraints of such a theory. Nevertheless, as we will demonstrate, backgrounds determined by the external supersymmetry constraints indeed lead to novel $7d$ field theories on curved spaces.

The curved Euclidean 7-branes that we wish to describe are solutions to type IIA$^*$ theory \cite{hull1, hull2, hull3}; Euclidean D$p$-branes will be referred to as E$(p+1)$-branes. Type IIA$^*$ theory corresponds to the timelike T-dual of type IIB, and although it has a Lorentzian metric, its RR-fluxes are imaginary and hence the kinetic terms in the action come with the `wrong' sign. An alternative way to derive the IIA$^*$ theory is to start with a complexified ten-dimensional action invariant under complexified supersymmetry and to deduce that, in addition to the usual reality constraints leading to type IIA, there is another set of reality constraints that one might impose instead \cite{bergshoeff}; in particular, the unusual signs are
\eq{
F^* = - F~, \qquad \ve^c =  \G_{(10)} \ve ~.
}
for the RR-fluxes $F$ and the Killing spinors $\ve = (\ve_1, \ve_2)$.
Our starting point will be to deduce the supersymmetry conditions of E7-branes wrapping curved spaces, which are obtained by imposing both the IIA$^*$ (closed string) supersymmetry as well as the worldvolume (open string) supersymmetry constraints.

The closed string supersymmetry constraints  are given by
\eq{\label{closed}
D_M \ve \equiv \left( \nabla_M + \frac14 \slashed{H}_M \pcal \right) \ve
- \frac18 e^\phi \left(F_0 \G_M  + \slashed{F_2} \pcal \G_M  + \slashed{F_4}\G_M  \right) \ve &= 0\\
D \ve \equiv \left( \slashed{\p} \phi + \frac12 \slashed{H} \pcal \right) \ve
- \frac18 e^\phi \left(10 F_0  - 6 \slashed{F_2} \pcal + 2 \slashed{F_4}  \right) \ve &= 0 ~.
}
These follow from the vanishing of the IIA$^*$ gravitino and dilatino variations.

The open string supersymmetry constraint is given by
\eq{\label{open}
\G \ve = \ve ~,
}
which is the necessary requirement for a brane to not break supersymmetry of a IIA$^*$ background. The Euclidean kappa-symmetry operator for E7-branes is given by
\eq{\label{kappa}
\G  = \frac{1}{7!} i \e^{\m\n\r\s\kappa\l \tau} \G_{\m\n\r\s\kappa\l \tau}  = i \pcal_1~,
}
with the second equality following from the $Spin(1,9) \rightarrow Spin(7) \times Spin(1,2)$ decomposition of the gamma-matrices given in appendix \eqref{gammadecomp}, where one can also find our definitions for $\pcal$, $\pcal_1$. Here, we have explicitly set the worldvolume flux $\fcal=0$ to substantially simplify matters. After decomposition of the Killing spinors and taking into account the reality constraints, the open supersymmetry constraint is solved by
\eq{\label{ks}
\ve_1 &=    e^{-\frac12 X(x,y)}  \chi_j(x) \otimes \zeta_j(y)  \otimes \left( \begin{array}{c} 1 \\ 0 \end{array} \right) \\
\ve_2 &=    e^{-\frac12 X(x,y)}  \chi_j(x) \otimes \zeta_j(y)  \otimes \left( \begin{array}{c} 0 \\ -i \end{array} \right)~.
}
Here, $\chi_j$ are Majorana spinors of $Spin(7)$ and $\zeta_j$ are Majorana spinors of $Spin(1,2)$, and the index $j$ runs over the number of independent internal spinors $\z_j$ (i.e., at most 2). We will assume a local decomposition of the ten-dimensional spacetime $M_{10} = M_7 \times M_3$ and use local coordinates $x^\m$ on the space $M_7$ wrapped by the E7-brane, and $y^a$ on $M_3$.

The supersymmetry constraints \eqref{closed}, \eqref{open} can be rewritten in a pair of sets with a more suggestive interpretation.
The first set of constraints is given by
\eq{
\frac12 \{D_\m, \G\} \ve = 0~, \qquad \frac12 \{D, \G \} \ve = 0 ~, \qquad \frac12 [D_a, \G ] \ve = 0 ~.
}
These equations are interpreted as determining the supersymmetry of the brane. They are given by
\begin{subequations}\label{L}
\begin{align}
\Big( \nabla_\m^{(7)} + \frac14 \o_{\m ab} \cg^{ab} + \frac14 i H_{\m\n a} \cg^a \g^\n
- \frac18 e^\phi F_0 \g_\m \nn\\
- \frac18 i e^\phi F_{\n a}\cg^a \g^\n \g_\m
- \frac18 e^\phi \left(\frac{1}{4!} F_{\n\r\s\l} \g^{\n\r\s\l} \g_\m + \frac14 F_{\n\r ab} \cg^{ab} \g^{\n\r} \g_\m  \right) \Big) \ve &= 0\\
\Big( \p_\m \phi \g^\m + \frac12 i \left( \frac12 H_{\m\n a} \cg^a \g^{\m\n} - H_{089} \right)
- \frac54 e^\phi F_0 \nn\\
+ \frac34 i e^\phi F_{\m a} \cg^a \g^\m
- \frac14 e^\phi \left( \frac{1}{4!} F_{\m\n\r\s} \g^{\m\n\r\s} + \frac14 F_{\m\n ab} \cg^{ab}  \g^{\m\n} \right) \Big)\ve &= 0 \\
\Big( \frac12 \o_{ab\m} \cg^b \g^\m + \frac18 i \left(H_{abc} \cg^{bc} + H_{\m\n a} \g^{\m \n} \right)
- \frac18   e^\phi F_0 \cg_a \nn\\
- \frac18 i e^\phi F_{\m b} \cg^b \cg_a \g^\m
- \frac18   e^\phi \left( \frac{1}{4!} F_{\m\n\r\s}  \cg_a \g^{\m\n\r\s} + \frac14 F_{\m\n bc}\cg^{bc} \cg_a  \g^{\m\n}  \right) \Big) \ve &= 0~.
\end{align}
\end{subequations}
The second set of equations is given by
\eq{
\frac12 [D_\m, \G] \ve = 0~, \qquad \frac12 [D, \G]\ve = 0 ~, \qquad \frac12 \{D_a, \G \} \ve = 0 ~,
}
and is interpreted as the supersymmetry of the string background in which the background is embedded. These are given by
\begin{subequations}\label{R}
\begin{align}
\Big(- \frac12 \o_{\m\n a} \g^\n \cg^a + \frac18 i \left( H_{\m\n\r} \g^{\n\r} + H_{\m ab} \cg^{ab} \right)
- \frac{1}{16} i e^\phi \left( F_{ab} \cg^{ab} + F_{\n\r} \g^{\n\r} \right) \g_\m \nn \\
+ \frac{1}{8}   e^\phi \left( F_{089\n} \g^{\n} + \frac{1}{3!} F_{\n\r\s a} \cg^a \g^{\n\r\s} \right) \g_\m \Big) \ve &= 0 \\
\Big( \p_a \phi \cg^a - \frac14 i \left( \frac13 H_{\m\n\r} \g^{\m\n\r} + H_{\m ab} \cg^{ab} \g^\m\right)
+ \frac38 i e^\phi \left( F_{\m\n} \g^{\m\n} + F_{ab} \cg^{ab} \right) \nn \\
+ \frac{1}{4} e^\phi \left( F_{089\m} \g^\m + \frac{1}{3!} F_{\m\n\r a} \cg^a \g^{\m\n\r} \right)\Big) \ve &= 0 \label{R2}\\
\Big(\nabla_a^{(3)} + \frac14 \o_{a\m\n} \g^{\m\n} - \frac14 i H_{\m ab}\cg^b  \g^\m
+ \frac{1}{16} i e^\phi \left( F_{\m\n} \g^{\m\n} + F_{bc} \cg^{bc} \right) \cg_a \nn \\
- \frac18 e^\phi \left( F_{089\m} \g^\m + \frac{1}{3!} F_{\m\n\r b} \cg^b \g^{\m\n\r}\right) \cg_a \Big) \ve &= 0~.
\end{align}
\end{subequations}
We will refer to \eqref{L} as the external supersymmetry constraints and \eqref{R} as the internal constraints, since from the point of view of the gauge theory on the brane, \eqref{R} dictates the (non-compact) space which determines the R-symmetry.\footnote{
Technically, we expect only the contraction of $\cg^a$ of the last external supersymmetry constraint to play a role in determining the supersymmetry constraints of the brane itself. Since we intend to solve all equations, the precise identification of external versus internal will play no role in what follows. }

Let us examine the various terms in \eqref{L} in terms of representations of the supersymmetry algebra. The Lorentzian supersymmetry algebra in $d=7$ comes in two types: maximal supersymmetry with 32 supercharges or half-maximal supersymmetry with 16 supercharges. Less supercharges is not possible since $Spin(1,6)$ does not admit 8 component Majorana spinors. The representations can be decomposed in terms of the product of the isotropy subgroup of the Lorentz group for lightlike vectors, $SO(5) \subset SO(1,6)$, and the R-symmetry group $SO(3) \simeq SU(2)$.\footnote{We will ignore subtleties about double covers.} In the Euclidean case, the R-symmetry group becomes $SO(1,2) \simeq SU(1,1)$. On the other hand, the Lorentz group is now $SO(7)$ which does not admit lightlike representations. We will forego this subtlety and identify representations as if they were representations of $SO(5) \times SU(2)$.

The (bosonic) fields in \eqref{L} can be identified the following representations of $SO(5) \times SU(2)$:
\begin{center}
{\renewcommand{\arraystretch}{1.2}
\begin{tabular}{ll|r}
Fields && $SO(5) \times SU(2)$ Representation \\ \hline
$g_{\m\n}$                         &  & ${\bf (14, 1)}$ \\
$F_{\m\n\r\s}$                     &$\sim \p_{[\m} C_{\n\r\s]}$ & ${\bf (10,1)}$\\
$F_{\m\n a b}$                     &$\sim \p_\m \e_{abc} C_\n^{\phantom{\n}c} $  & ${\bf (5,3)}$ \\
$H_{\m\n}^{\phantom{\m\n}a}$       &$\sim \p_\m B_\n^{\phantom{\n}a} $ & ${\bf (5,3)}$ \\
$\o_{ab\m}$                        &$\sim  g_{ab} \p_\m A $     & ${\bf (1,5)}$ \\
$F_{\m}^{\phantom{\m}a}$           &$\sim \p_\m C^a$ & ${\bf (1,3)}$ \\
$H_{abc}$                          &$\sim h \e_{abc} $          & ${\bf (1,1)}$ \\
$\phi$                             &       & ${\bf (1,1)}$
\\\hline
\end{tabular}
}
\end{center}
$F_0$ should be considered as a cosmological constant with no degrees of freedom, while $\o_{\m ab}$ is locally pure gauge, and hence neither are included.
The Lorentzian $d=7$ supersymmetry algebra with 16 supercharges admits the following three types of multiplets \cite{strathdee}:
\begin{itemize}
\item 40+40 gravity multiplet: $\{ {\bf (14,1), (10,1), (1,1), (5,3)} ~| ~{\bf (16,2), (4,2)} \}$
\item 24+24 $SU(2)$-vector multiplet: $\{ {\bf (1,5), (1,1), (1,3), (5,3)} ~|~ {\bf (4,4), (4,2)} \}$
\item 8+8 vector multiplet: $ \{ {\bf (5,1), (1,3)} ~|~ {\bf (4,2)} \}$
\end{itemize}
All bosonic components of a gravitational multiplet and an $SU(2)$-vector multiplet are accounted for.\footnote{There is some ambiguity in identifying which ${\bf (5,3) }$ field belongs to which multiplet and similarly for the two ${\bf (1,1)}$ fields, but this is irrelevant for our purpose.} We thus conclude that the external supersymmetry equations \eqref{L} correspond to the supersymmetry variations of the fermionic fields of a Euclidean $7d$, $\ncal =1$ off-shell supergravity coupled to a single $SU(2)$-vector multiplet.

\section{Backgrounds on $G_2$-structure manifolds}\label{g2sol}
In this section, we will construct classes of solutions to the external and internal supersymmetry constraints \eqref{L}, \eqref{R}, with fluxes that satisfy the Bianchi identities.
Strictly speaking, from the point of view of coupling the field theory to a background, neither the internal constraints nor the IIA$^*$ equations of motion need be satisfied.
Nevertheless, we will solve the internal constraints as well, as a stepping stone to finding
E7-branes solutions in IIA$^*$ theory satisfying the IIA$^*$ equations of motions; as we will show, such solutions would require a non-trivial E7-brane worldvolume flux which is associated to non-linear supersymmetry \cite{mpt} and is beyond the scope of this paper.

Our primary tool in solving the supersymmetry constraints will be a $G_2$-structure, see for example \cite{cs},\cite{fkms} for reviews and appendix \ref{g2} for our conventions. On manifolds in $d=7$, a unit-norm nowhere-vanishing (Majorana) spinor $\chi$ of $Spin(7)$ is equivalent to a $G_2$-structure, defined by an associative three-form $\vf$ and its dual $\psi = \star_7 \vf$.
Given a $G_2$-structure, we may decompose all fields in terms of irreducible representations of $G_2$, and the connection acting on the Killing spinor can be expressed algebraicly in terms of torsion classes via \eqref{nablachi}, thus allowing us to solve all equations algebraicly. In this way, we obtain two classes of backgrounds: one on manifolds equipped with a co-closed $G_2$-structure, the other on conformally $G_2$-holonomy manifolds.
Since there is no additional topological constraint to the existence of co-closed $G_2$-structures \cite{cn}, we conclude that any spin 7-manifold admits solutions to the supersymmetry constraints.

The amount of supersymmetry preserved by such solutions is determined
by the number of inequivalent Killing spinors leading to equivalent solutions. Due to the Killing spinor decomposition \eqref{ks}, the number of ten-dimensional supercharges is the product of the number of external Killing spinors $\chi_j$ (8 or less) and internal Killing spinors $\z_j$ (2 or less). As we will see, for non-flat spaces our solutions have $F_4$ proportional to the $G_2$-structure itself; since no two linearly independent spinors lead to the same $G_2$-structure, our solutions will preserve two supercharges. This is less than the `minimum' number of supercharges constituting the flat supersymmetry algebra (i.e., 8 supercharges for $Spin(7)$). This is not uncommon in Euclidean field theories, but does limit computational control.\footnote{As an example, consider that the $d=4$, $\ncal =1$ supersymmetry algebra has four supercharges, whereas field theories can be put on $K3$ preserving only two supercharges.}
We will come back to this point later on.

\subsection{External supersymmetry}
Our primary concern will be the external supersymmetry constraints \eqref{L}, as these dictate the geometry of the brane.
We decompose all fields in terms of $G_2$-structure representation, which immediately leads to the following observation. The torsion classes are uncharged under the R-symmetry (i.e., they have no internal space indices $a,b,..$ taking value in $0,8,9$). However, there are no uncharged fluxes taking value in the  ${\bf 14}$ representation, hence the internal Lorentz symmetry remaining unbroken immediately imposes $\tau_2 = 0$.

We will focus on the case where the R-symmetry is not broken for simplicity and hence we will set the profiles of fields in the ${\bf 3}$ of $SO(1,2)$ to zero by hand,
\eq{
H_{\m\n a} = \o_{\m a b} = F_{\m a} = F_{\m\n a b} = 0~.
}
Note that due to the decomposition into the irreducible representations, non-trivial profiles for these fields would have no effect on the results obtained for ${\bf 1}$ and ${\bf 27}$ representations which will play a crucial role in our field theory discussion in section \ref{ft}.
We will set the metric to
\eq{\label{g}
ds_{10} = e^{2A(x,y)} ds_7^2(x) + e^{2B(x,y)} ds_3^2(y)~.
}
Next, we decompose the remaining fluxes in terms of $G_2$-representations via
\eq{\label{f4decomp}
F_{\m\n\r\s} &= i \left(4 \vf_{[\m\n\r} f_{\s]} + f \psi_{\m\n\r\s} + f_{[\m}^{\phantom{[\m}\l} \psi_{\n\r\s]\l} \right)~, \qquad
H_{abc} = h \e_{abc}
}
and use the identities \eqref{g22} to rewrite all terms in $\eqref{L}$ in terms of the spinorial basis $(\chi, \g_\m \chi)$, and finally use the crucial \eqref{nablachi} to express the connection acting on $\ve$ in terms of torsion classes. As a result, $\eqref{L}$ decomposes into a set of algebraic equations that can be solved representation by representation. We thus obtain the following solution to the external supersymmetry constraints:
\eq{
F_0 &= 0 \\
e^{A+ \phi} f &= \frac27 e^A h =  4 \tau_0 \\
e^{A+ \phi} f_\m &=  \p_\m \phi =  \p_\m A =  \tau_{1\m} \\
e^{A+ \phi} f_{\m\n} &=  \tau_{ \m\n}  ~.
}
The Kiling spinor $\ve$ is given by
\eq{
\ve = e^{-\frac12 X(x, y)} \chi(x)  \otimes \zeta(y) \otimes \left(\begin{array}{c} 1 \\ -i \end{array} \right) \\
}
with $X$ constrained to satisfy $\p_\m X = \p_\m A$ and $\chi$ the Majorana spinor of unit norm that defines the $G_2$-structure. All indices refer to unwarped metrics.

\subsection{Internal supersymmetry}
Next, let us consider the internal supersymmetry constraints \eqref{R}. These dictate the embedding conditions for a supersymmetric brane into a supersymmetric type IIA$^*$ background. In principle, various solutions may exist which can be obtained in a similar manner as the external supersymmetry solutions by making use of the $G$-structure defined by a single Majorana spinor of $Spin(1,2)$. However, we are not particularly concerned with generality in this respect and so will restrict ourselves to a simple solution, namely warped flat space. We write the metric as the cone over dS$_2$:
\eq{
ds_3^2 &=  ds^2(\rbb^{1,2}) = \d r^2 + r^2 (- \d t^2 + \cosh^2 t \d \vt^2)~.
}
Our ansatz for the fluxes is to take\footnote{Note that the index $r$ refers to the radial coordinate $y^9 = r$ and is not a summation index.}
\eq{
H_{\m\n\r} &= F_{\m\n} = F_{089\m} = H_{\m a b} = F_{\m\n\r a} = 0 \\
F_{ a b} &= - i q r^{-2} \e_{r ab} ~.
}
In particular, note that $H_{\m\n\r} = 0$ is necessary due to the fact that we are using \eqref{kappa} where we have set the worldvolume flux of the E7-brane to zero. In addition, we also take the ansatz
\eq{
A (x,y) &= A_{\text{int}}(r) + \Delta (x) \\
B(x,y) &= \frac14 \log Z (r) + B_{\text{ext}} (x) \\
\phi(x, y) &= \phi_{\text{int}}(r) + \phi_{\text{ext}}(x)~.
}
We then obtain the following solution to the internal supersymmetry constraints \eqref{R}:
\eq{
\frac14 \p_r \log Z &= - \p_r \Delta =  - \frac13 \p_r \phi_{\text{int}} \\
 &= - \frac14 e^{\phi-B} q r^{-2} ~,
}
with the additional constraint on the Killing spinor that $\p_r X = - \p_r B = + \p_r A$. Note that the sign is relevant: since we do not have $\p_M X = \p_M A$, this imposes an additional constraint on the warp factors $A$, $B$. Taking into account the external supersymmetry constraints, the solution to the differential equations is given by
\eq{
Z = k + \frac{q}{r}~
}
with $k, q \in \rbb$, as well as $\phi_{\text{int}}(r) = - 3  A_{\text{int}} = \frac34 \log Z (r)$.

\subsection{Background constraints}\label{sec:g2sol}
Let us now combine the constraints from the previous two subsections. Using the ansatz in the last paragraph of the previous section, $A, B$ and $\phi$ are fixed. We then obtain the following solution to the supersymmetry constraints \eqref{L}, \eqref{R}:
\eq{\label{thesolution}
ds_{10}^2 &=  e^{2 \Delta(x)} \left( \frac{1}{\sqrt{Z}} ds^2_7 (x) + \sqrt{Z} ds^2 (\rbb^{1,2}) \right) \\
Z(r)     &= k + \frac{q}{r}\\
\phi &= - \frac34 \log Z + \Delta \\
H    &=  14 \tau_0 e^{2 \Delta} Z  \vol_3 \\
F_0  &= 0 \\
F_2  &= - i q \vol_2 \\
F_4  &=  i e^{2 \Delta} \left(  4 \tau_0 \psi - \tau_1 \wedge \vf  + \tau_3 \right)\\
&\\
\tau_1  &=   \d \Delta \\
\tau_2  &= 0 ~,
}
with Killing spinor
\eq{\label{kssol}
\ve =  Z^{-1/8} e^{-\frac12 \Delta(x)} \chi(x)  \otimes \zeta(y) \otimes \left(\begin{array}{c} 1 \\ -i \end{array} \right)~.
}
Here, $\vol_2 =  \cosh t \d t \wedge \d \vt$ is the volume of dS$_2$ and $\vol_3 =  r^2 \cosh t \d r \wedge \d t \wedge \d \vt$ is the volume of $\rbb^{1,2}$.
Imposing the Bianchi identities, we find two classes of geometries that solve the supersymmetry constraints:
\eq{
\text{I: }& \qquad \tau_1 = \tau_2 =  \d \tau_0 = \d \tau_3 = 0 \\
\text{II: }& \qquad \tau_0 = \tau_2 =  \tau_3 = \tau_1 - \d \Delta = 0~.
}
In the first case, the $G_2$-structure is co-closed, and by rescaling the coordinates one can take $\Delta = 0$ without loss of generality.
In the latter case $(M_7, ds^2_7)$ has conformal $G_2$-holonomy.

Although not relevant from a field theory point of view, let us investigate the equations of motions. We adapt the integrability theorem, first derived for IIA in \cite{lt}, to the case of IIA$^*$ at hand. Examining the computations in \cite{lmmt} (see also \cite{pt}), it follows that the IIA integrability theorem also holds for type IIA$^*$, despite complexification of the fields or any potential breaking of Lorentz symmetry. Therefore, all NSNS field equations are satisfied, provided that the generalized Bianchi identities for the democratic RR-fluxes hold and the mixed space-time components of the NSNS field equations are satisfied. These conditions are violated precisely by the equation of motion of $F_4$ (or in terms of the democratic formalism, the Bianchi identity of $F_6$) if and only if any torsion class is non-vanishing.

\section{Examples of backgrounds}
In this section, we give a number of explicit examples of solutions in the class described by \eqref{thesolution}. We emphasize again that, although these do not generally satisfy the equations of motion, these solutions form backgrounds to which field theories can be coupled supersymmetrically.

\subsection{Ricci-flat space}
As a first example, we consider flat E7-branes with no torsion. The solution to the supersymmetry constraints \eqref{thesolution} then reduces to
\eq{
ds_{10}^2 &=  \frac{1}{\sqrt{Z}} g_{\m\n} \d x^\m \d x^\n + \sqrt{Z} \left( \d r^2 +  r^2 ( - \d t^2 + \cosh t \d \vt^2 ) \right) \\
Z         &= k + \frac{q}{r} \\
e^{2\phi} &= Z^{-\frac32} \\
F_2       &= - i q \cosh t \d t \wedge \d \vt ~.
}
The Bianchi identities are manifestly satisfied, as well as the equations of motion. Since $\nabla_\m \chi = 0$, it follows that $g_{\m\n}$ must be a Ricci-flat metric. The case $g_{\m\n} = \delta_{\m\n}$ corresponds to the expected solution where the E7-brane wraps $\rbb^7$ (or $\rbb^n \times T^{7-n}$) \cite{hull2}. This solution preserves 16 supercharges. It should be noted however that various other solutions exist; generic $G_2$-holonomy manifolds preserve 2 supercharges, $K3 \times T^3$ preserves 8 supercharges, and so on.

\subsection{Weak $G_2$-holonomy}\label{s7}
As a second example, we consider manifolds admitting a Killing spinor satisfying
\eq{
\nabla_\m \chi = \frac12 i \g_\m \chi~,
}
i.e., $\tau_0 = 1$ and all other torsion classes vanishing. In this case, \eqref{thesolution} reduces to
\eq{\label{scalartorsion}
ds_{10}^2 &=  e^{\frac 23  \phi}  ds^2(M_7)  + e^{-\frac23 \phi} ds^2 (\rbb^{1,2}) \\
e^{2\phi} &=  \left( k + \frac{q}{r} \right)^{-\frac32} \\
H         &=  14 e^{ - \frac43   \phi} \vol_3 \\
F_2       &=  - i q \vol_2 \\
F_4       &=  4 i  \psi ~.
}
Manifolds equipped with a $G_2$-structure with purely scalar torsion are known as nearly-parallel, or equivalently, weak $G_2$-holonomy spaces.
In particular, as determined in \cite{bar} (see also \cite{prz}), generic Sasaki-Einstein manifolds admit 2 such Killing spinors, generic tri-Sasakians admit 3 and the round $S^7$ admits 8. Due to the four-form flux $F_4 \sim \psi$, the solution is invariant under $G_2 \times SO(1,2)$ or a subgroup thereof.

Let us consider $S^7$ in particular. On $S^7$, we consider the metric
\eq{
ds^2_\l(S^7) = \d \m^2 + \frac14 \sin^2 \m (\s_i - \ti{\s}_i)^2 + \frac14 \l^2 \left[ (\s_i + \ti{\s}_i)^2 + \cos \m (\s_i - \ti{\s}_i)^2 \right]~.
}
Here, $\s_i, \ti{\s}_i$ are two pairs of $SU(2)$ left-invariant forms, each parametrizing an $S^3$. There are two values of the parameter $\l$ for which this metric is Einstein: $\l^2 =1$, which yields the standard round metric, and $\l^2 = \frac15$ \cite{jensen} (see also \cite{dnp}). The latter case corresponds to a squashing of the $S^3$ fiber over the base space $S^4$. The $S^3$-squashed $S^7$ admits a single Killing spinor.
The round $S^7$ admits 8 Killing spinors, which transform as the {\bf 8} of $SO(8)$. This breaks as ${\bf 8} \rightarrow {\bf 1} \oplus {\bf 7}$ under $Spin(7) \rightarrow G_2$. In particular, the ${\bf 7}$ can be thought of as the component perpendicular to ${\bf 14}$ under the breaking of $Spin(7)\rightarrow G_2$,
with ${\bf 14}$ the adjoint which leaves $G_2$ invariant. In other words, given two linearly independent Majorana spinors $\chi_{1,2}$, both will lead to different $G_2$-structures and thus, $F_4$ in \eqref{scalartorsion} cannot remain invariant.\footnote{
 This can be made more explicit by comparing $\vf^{(2)}$ with $\vf^{(1)}$ for $\chi_2 = a \chi_1 + i b_\m \g^\m \chi_1$.}
Therefore, the solution preserves 2 supercharges after taking into account the R-symmetry. This is somewhat analogous to what occurs for M-theory solutions on $S^7$ where an internal $F_4$ is turned on which breaks supersymmetry \cite{englert}, although the difference in properties of spinors of $Spin(2,9)$ have a non-neglible effect on this interpretation when uplifting our solution to M$^*$-theory.\footnote{See also \cite{ranjbar}.}

It is interesting to compare our results for branes wrapping $S^7$ to those of \cite{bbg}. The BBG-solution can be viewed as a background of $8d$ supergravity, where the metric is a domainwall containing an $S^7$ factor. This background is then uplifted to type IIA$^*$ and is expected to be holographically dual to the known $7d$ SYM theory on $S^7$ given by \eqref{przaction}, \eqref{przsusy}. It comes with $H \sim \tau_0 \vol_3$, but has $F_4 = 0$. As a result, it does not break any supersymmetry. It also satisfies the equations of motion explicitly. On the other hand, it does not satisfy kappa-symmetry and should thus be interpreted as the near-horizon limit of an E7-brane. Since our results, obtained by imposing kappa-symmetry from the start, require $F_4 \sim \psi$, we conclude that any supersymmetric brane satisfying the equations of motion with the BBG-solution as a near-horizon cannot preserve the $SO(8)$-isometry; the near-horizon limit leads to symmetry enhancement.

\subsection{$U(1)$-squashed $S^7$}
A more intricate example is the case where both $\tau_0 \neq 0$, $\tau_3 \neq 0$. This occurs for $S^7$ where it is the $S^1$ Hopf fibre that is squashed. The metric for this case is given by \cite{lpr}
\eq{
ds^2(S_{\ti{l}}^7) = l^2 ds^2 (\cbb P^3) + \tilde{l}^2 \left( d \t + \acal \right)^2~.
}
Here, $l, \ti{l} \in \rbb$ and $\acal$ is the fiber connection, which satisfies $\d \acal = 2 J$, with $J$ the K\"{a}hler form of $\cbb P^3$.
As shown in \cite{lpr}, this squashed $S^7$ admits a Dirac Killing spinor $\eta$ satisfying
\eq{
\nabla_\m \eta =  \frac{\a}{2l} i \g_\m \eta + 2 (\a^2 - 1) i \acal_\m \eta~,
}
but for our purposes, it is more convenient to work with Majorana spinors instead. By setting $\eta = \frac12 ( 1 + \tilde{\acal}_\m \g^\m) \chi$, it is found after some manipulation that
\eq{
\nabla_\m \chi = \frac12 i \left( \frac{\a}{l} + \frac47 \frac{l}{\a} (\a^2 -1) \right)  \g_\m \chi +  \frac{2 l}{\a} (\a^2 -1) i \left( \frac{\a^2}{l^2} \acal_{(\m}  \acal_{\n)} - \frac17 g_{\m\n} \right) \g^\n \chi~,
}
from which we read off the torsion classes
\eq{
\tau_0 &=  \frac{\a}{l}\left( 1 + \frac47 \frac{l^2}{\a^2} (\a^2 -1) \right)\\
\tau_{\m\n} &= - 16 \frac{ l}{\a} (\a^2 -1)  \left( \frac{\a^2}{l^2} \acal_{(\m}  \acal_{\n)} - \frac17 g_{\m\n} \right)~.
}
Thus, the $U(1)$-squashed $S^7$ can be considered as a deformation of the round $S^7$,
leading to a modification of the ${\bf 1}$ torsion and turning on a non-zero ${\bf 27}$ torsion.
A second Majorana Killing spinor with identical torsion classes is given by
\eq{
\chi_2 = - i \frac{\a}{l} \acal_\m \g^\m \chi~,
}
which is obtained after noting that $\tilde{\acal} = \a/ l \acal$ for consistency.
We therefore find the resulting background
\eq{
ds_{10}^2   &=  \frac{1}{\sqrt{Z}} ds^2 (S^7_{\ti{l}}) + \sqrt{Z} ds^2 (\rbb^{1,2}) \\
e^{2\phi} &=  \left( k + \frac{q}{r} \right)^{-\frac32} \\
H         &=  14 e^{ - \frac43   \phi} \vol_3 \\
F_2       &= - i q \vol_2 \\
F_4       &=  4  \frac{\a}{l} \left(1 + \frac{4 l^2}{7 \a^2} (\a^2 -1) \right) i  \psi + i {\tau}_3   ~,
}
with $\tau_{3\m\n\r\s} =   \tau_{[\m}^{\phantom{[\m}\l} \psi_{\n\r\s]\l}$ as per usual.

\subsection{$S^3 \times M_4$}\label{s3t4}
Next, we will construct solutions on $S^3 \times M_4$, where $M_4$ is any hyperk\"{a}hler manifold. For convenience, we will first consider $M_4 = T^4$ and then generalize to arbitrary hyperk\"{a}hlers. Our starting point will be the associative three-form $\vf$ and its dual $\psi$, from which we will deduce the Killing spinor.

We make use of the standard expression for the associative three-form $\vf$ in terms of vielbeine \cite{cs}:
\eq{\label{s3t4psi}
\vf &= e^{567} + \left(e^{13} - e^{24} \right) e^5 - \left(e^{14} + e^{23} \right) e^6 + \left(e^{12} + e^{34}\right) e^7 \\
\psi &= e^{1234} + \left( e^{13} - e^{24} \right)e^{67} + \left(e^{14} + e^{23} \right) e^{57} + \left(e^{12} + e^{34}\right) e^{56} ~.
}
Since both $S^3$ and $T^4$ admit an identity structure (i.e., they are parallelizable), these expressions are globally well-defined. We choose a frame $e^{1,2,3,4}$ on $T^4$ and $e^{5,6,7} = K^{1,2,3}$ on $S^3$, with $K^{1,2,3}$ the left-invariant forms satisfying the Maurer-Cartan equation
\eq{
\d K^j = \frac12 \e^{jkl} K^k \wedge K^l~.
}
It thus follows that
\eq{
\d \phi &= \left( e^{13} - e^{24} \right) K^{23} + \left(e^{14} + e^{23} \right) K^{13} + \left(e^{12} + e^{34}\right) K^{12}\\
\d \psi &= 0~,
}
from which we read off the torsion classes
\eq{\label{s3t4w}
\tau_0 &=  \frac{3}{14}~, \qquad \tau_1 = 0 ~, \qquad \tau_2 = 0 ~, \\
\tau_3 &= - \frac67 e^{1234} + \frac17 \left[  \left( e^{13} - e^{24} \right)K^{23} + \left(e^{14} + e^{23} \right) K^{13} + \left(e^{12} + e^{34}\right) K^{12} \right]~.
}
From the four-form $\tau_3$, it follows that the traceless symmetric two-tensor $\tau$ can be expressed in terms of the frame as
\eq{
\tau =  - \frac67 \left( e^a \otimes e^a \right) + \frac87 \left(K^j \otimes K^j \right)~.
}
Since the $G_2$ structure is co-closed, it follows that the E7-brane supersymmetry constraints are solved for $ds_7^2 =  ds^2(S^3) + ds^2(T^4)$. Explicitly, plugging the above into \eqref{thesolution} leads to the following background:
\eq{\label{s3t4sol}
ds_{10}^2 &= \frac{1}{\sqrt{Z}} \left( ds^2 (S^3) + ds^2(T^4)\right) + \sqrt{Z} ds^2 (\rbb^{1,2}) \\
Z(\r)     &= k + \frac{q}{\r}\\
H    &=  3  Z  \vol_3 \\
\phi &= - \frac34 \log Z \\
F_2  &= - i q \vol_2 \\
F_4  &= - i \left(  \frac67 \psi   + \tau_3 \right)~,
}
with $\psi$, $\tau_3$ as in \eqref{s3t4psi}, \eqref{s3t4w}.

More generally, let us introduce the two-forms
\eq{
\begin{alignedat}{6}
\a_1 &{}={}& (e^{13}& - e^{24} )~, &\qquad& \a_2 &{}={}& (e^{14} + e^{23} ) ~, &\qquad& \a_3 &{}={}& (e^{12} + e^{34} ) ~.
\end{alignedat}
}
We note that the above procedure can be repeated identically as long as $\a_{1,2,3}$ are globally well-defined closed two-forms, even in the absence of an identity structure. Therefore, identical solutions can be found on $S^3 \times M_4$ for any $M_4$ that is hyperk\"{a}hler,

\section{Gauge theory on curved 7-branes}\label{ft}
In this section, we will derive field theory actions on the curved 7-brane backgrounds obtained in section \eqref{g2sol}. The canonical example is SYM on $S^7$. This theory was first constructed in \cite{blau}, taken off-shell in \cite{fhy} using the flat space procedure of \cite{berk},\cite{evans}, and further investigated in \cite{mz},\cite{prz}. The action is composed of the standard flat space SYM action, in addition to a number of curvature terms which were added somewhat ad hoc. Following the logic of \cite{fs}, these terms should correspond to coupling the theory to a supersymmetric background determined by an off-shell supergravity which, as we have argued, should be reproducable by a string background. Indeed, we will show that these curvature terms can all be expressed as coupling to a background $H$-field.
Given our results in section \ref{g2sol}, we argue that the following should be true:
\begin{itemize}
\item In addition to coupling gauge theory on $S^7$ (or more generally, any weak $G_2$-holonomy manifold) to $H$, it should also be possible to couple gauge theory to a background $F_4$, whose profile is determined by \eqref{thesolution}.
\item Supersymmetric gauge theories should exist on Riemannian manifolds in the following two classes: co-closed $G_2$-structure spaces and conformally $G_2$-holonomy spaces.
\item In the case $\tau_3 \neq 0$, the coupling to $F_4$ is essential, since that is the only background field affected by such torsion.
\end{itemize}
We shall demonstrate these claims below for the first class of manifolds. Inspired by the DBI-action, we will construct a supersymmetric action on co-closed $G_2$-structure manifolds, where a coupling to $F_4$ (or more precisely, $C_3$) is made explicit. From a more geometric point of view, the novel feature of this action is the inclusion of what we will refer to as a ``$G_2$-Chern-Simons" term \cite{dt}, as well as fermionic counterterms such that two supercharges are preserved on-shell.\footnote{A supersymmetric contact structure Chern-Simons action in $7d$ was constructed in \cite{kz}. This is a different type of Chern-Simons-like theory from the one discussed here.} We will take one of these supercharges off-shell, which suffices to allow for localization.

\subsection{SYM on weak $G_2$-holonomy manifolds}
Our starting point is the on-shell $d= 10$, $\ncal = 1$ SYM action on flat space, given by
\eq{
S = \int d^{10} x \text{Tr}\left( \frac12  F^{MN} F_{MN} - \overline{\Psi} \G^M D_M \Psi \right) ~,
}
with $\overline{\Psi} = \Psi^T C^{-1}$ and
\all{
D_M \Psi = \nabla_M \Psi + [A_M, \Psi] ~, \quad D_M A_N = \nabla_M A_N + [A_M, A_N] ~, \quad F_{MN} = \p_M A_N - \p_N A_M + [A_M, A_N] ~.
}
The action is invariant under the supersymmetry transformations
\eq{
\delta_\ve A_M  &= \overline{\Psi} \G_M \ve \\
\delta_\ve \Psi &= \frac12 F_{MN} \G^{MN} \ve
}
with $\ve$ a commuting $Spin(1,9)$ Majorana-Weyl spinor satisfying $\nabla_M \ve = 0$. By dimensional reduction on the flat internal Lorentzian space, this leads to the action
\eq{
S = \int d^{7} x \text{Tr}\left( \frac12  F^{MN} F_{MN} - \overline{\Psi} \G^M D_M \Psi \right) ~,
}
where we still use ten-dimensional indices to keep notation compact. The gauge field $A_M$ decomposes into $(A_\m (x), \phi_a(x))$ such that $F^2$ leads to kinetic terms for both the gauge field and scalar field, as well as a $\phi^4$ potential for the scalars. The gaugino $\Psi$ is an anti-commuting $Spin(1,9)$ Majorana-Weyl spinor. Gauge group indices have been suppressed.

Next, let us review the how to generalize the $7d$ Euclidean action, on the one hand to manifolds admitting Killing spinors with scalar torsion \cite{blau}, on the other to simultaneously allow for a partial off-shell formulation with up to nine supercharges off-shell \cite{berk},\cite{fhy}. This includes the canonical examples on $S^7$ and on Sasaki-Einstein manifolds. We will mostly follow the notation of \cite{prz}.
We decompose
\footnote{
There is a slight subtlety for the Killing spinor $\ve$: previously, we deduced that although kappa-symmetry reduces the number of supercharges to 16, $\ve$ is not chiral. Here, we will instead make use of a chiral spinor. }
\eq{\label{ftks}
\ve = \chi \otimes \z \otimes \left(\begin{array}{c} 1 \\ 0 \end{array} \right)  ~.
}
Using $\nabla_\m \chi =  \frac12 i \tau_0 \g_\m \chi$,
we obtain
\eq{\label{t0ks}
\nabla_\m \ve =   \frac12 \tau_0 \G_\m \L \ve~, \qquad \L \equiv \G_{089} = \pcal_2~.
}
The off-shell action on a weak $G_2$-holonomy manifold is then given by
\eq{\label{przaction}
S = \int d^7 x \sqrt{ g_{(7)}}\left( \frac12  F^{MN} F_{MN} - \overline{\Psi} \left(\G^M D_M - \frac32 \tau_0 \L\right) \Psi  + 8 \tau_0^2 \phi^a \phi_a  + 2 \tau_0 [\phi^a, \phi^b] \phi^c \e_{abc} - K^j K_j \right)~.
}
The action is invariant under the following supersymmetry transformations :
\eq{\label{przsusy}
\delta_\ve A_M  &= \overline{\Psi} \G_M \ve \\
\delta_\ve \Psi &= \frac12 F_{MN} \G^{MN} \ve + \frac87 \G^{\m a } \phi_a  \nabla_\m \ve + K^j \nu_j \\
\delta_\ve K^j  &= - \ol{\n^j} (\G^M D_M - \frac32 \tau_0 \L) \Psi ~.
}
The number of supercharges preserved is given by the number of Killing spinors admitted by the underlying manifold $M_7$ times the number of generators for the $SU(1,1)$ R-symmetry; for example, 16 for $S^7$, 4 for general Sasaki-Einstein manifolds, 2 for arbitrary weak $G_2$-holonomy manifolds. The  fields $K_j$ are 7 auxiliary scalar fields which take at most $9$ supercharges off-shell. The fact that $j \in \{1,...,7\}$ is incidental and unrelated to the fact that we are working in $d=7$.
The supersymmetry transformations of $K_j$ are given in terms of $\n_j$, a set of commuting pure spinors satisfying
\eq{\label{nuidentities}
\overline{\ve} \G_M \n_j &= 0 \\
\overline{\n_i} \G_M \n_j &= \delta_{ij}  v_M \\
\n^j \otimes \n_j + \ve \otimes \ve &= \frac12 v^M \G_M~.
}
with $v_M = \ol{\ve} \G_M \ve$. They are determined only up to $SO(7)$ rotations.

Taking into account \eqref{thesolution}, we now observe that the curvature terms in the action may be reformulated as the coupling to background $H$-fields. Specifically, setting $\Delta = Z = 1$ and making use of
\eq{
H = 14 \tau_0 \vol_3~,
}
we see that \eqref{przaction} can be rewritten as
\al{\label{przfluxes}
S =& \int d^7 x \sqrt{ g_{(7)}} ~\Tr \left( \frac12  F^{MN} F_{MN} - \overline{\Psi} \G^M D_M  \Psi - K^j K_j \right)\\
 +& \int d^7 x \sqrt{ g_{(7)}} ~\Tr \left(- \frac{1}{56} H^{MNP} \ol{\Psi} \G_{MNP} \Psi
 + \frac{1}{7} H_{MNP} A^M F^{NP}
 - \frac{1}{49} H_{MPQ} H^{NPQ} A_M A^N  \right)~.\nn
}
Note that crucially, $H$ is independent of the $G_2$-structure and hence does not break supersymmetry.

\subsection{Supersymmetric $G_2$-Chern-Simons}
In order to examine gauge theory on manifolds with additional non-trivial intrinsic torsion $\tau_3 \neq 0$, we wish to write down a supersymmetric action coupled to $C_3$. We will first do this for the simpler case on weak $G_2$-holonomy manifolds, which we will generalize to arbitrary co-closed $G_2$-structure manifolds afterwards. In the former case, one has $\d \vf = 4 \tau_0 \psi$. The brane solution \eqref{thesolution} yields the conclusion that
\eq{
-i F_4 =  \d \vf = 4 \tau_0 \psi \implies C_3 = i \vf ~.
}
We thus deduce from the WZ-part of the E7-brane action that coupling the field theory to $F_4$ should induce a term
\eq{
S_{F_4} &= - i \int C_3 \wedge F \wedge F = \int \vf \wedge F \wedge F \\
 &=  4 \tau_0 \int \psi \wedge \left( A \wedge \d A + \frac23 A \wedge A \wedge A\right) ~.
}
This is the $G_2$-Chern-Simons term \cite{dt}, which has appeared in various contexts in string theory \cite{bmss} \cite{afls} \cite{bln}, albeit generally for $G_2$-holonomy manifolds. In $3d$, the Chern-Simons action
\eq{
S_{CS} = \frac{k}{8 \pi^2} \int_{M_3} \Tr\left( A \wedge \d A + \frac23 A \wedge A \wedge A\right)
}
 is not gauge invariant and leads to quantization of the CS-level $k$. On the other hand, in $4d$, the $\theta$-action
\eq{\label{theta}
S_\t = \frac{\t}{16 \pi^2} \int_{M_4} F \wedge F
}
is gauge invariant. We see that in $7d$ on weak $G_2$-holonomy manifolds, up to a total derivative, the terms are equivalent. The gauge invariance is manifest, and so no quantization occurs. Similarly to the $\t$-action, the $G_2$-Chern-Simons action is intimately related with instantons. The equation of motion of \eqref{theta} is that the connection $A$ should be flat, while in the presence of a kinetic term, the equation of motion implies (anti-)selfduality and primitivity of $F$. In $7d$, the analogy of the flat connection is that the ${\bf 7}$ component of the field strength is trivial and hence the field strength takes value in the ${\bf 14}$ of the canonical $G_2$-structure associated to the metric. The analogous statement to selfduality is given by $\star_7 F = \vf \wedge F$. Generally, the occurrence of instantons in $p+1$ dimensions is associated to euclideanized D$p$-D$(p-4)$ systems \cite{witten},\cite{douglas}. In our case, no E3-brane is present, as can be confirmed by noting that the metric in \eqref{thesolution} preserves $SO(7)$ Lorentz symmetry. Instead, these instantons are sourced by the scalar curvature $\tau_0$.

We wish to construct a supersymmetric version of the $G_2$-Chern-Simons action coupled to Yang-Mills. For non-trivial $\tau_0$, a second realization of the supersymmetry algebra can be found in addition to \eqref{przaction}, \eqref{przsusy}, as shown in appendix \ref{ftcalc}.
The action
\al{\label{t0offshellaction}
S  = & \int d^7 x \sqrt{ g_{(7)}}~ \Tr \left( \frac12  F^{MN} F_{MN} - \overline{\Psi} \G^M D_M \Psi - K^j K_j \right) \nn\\
& +\tau_0 \int d^{7} x \sqrt{ g_{(7)}} ~ \Tr \left( -  \frac32 \overline{\Psi} \Lambda \Psi  -2 \e_{abc} F^{ab} \phi^c + 8  \tau_0 \phi^a \phi_a \right)\\
            & -  \int   4 \tau_0 \psi  \wedge \Tr\left( A \wedge \d A + \frac23 A \wedge A \wedge A\right)
              - \int d^7 x \sqrt{ g_{(7)}}~ \Tr  \left( \frac{1}{4!}  \tau_0 \psi^{\m\n\r\s} \overline{\Psi} \Lambda \G_{\m\n\r\s} \Psi \right)~.\nn
}
is invariant under the following on-shell supersymmetry transformations:
\eq{\label{t0offshellsusy}
\delta_\ve A_M &= \ol{\Psi} \G_M \ve \\
\delta_\ve \Psi &= \frac12 F_{MN} \G^{MN} \ve - \frac87 \G^{\m\a} \phi_a \nabla_\m \ve - 8 A^\m \nabla_\m \ve + \n^j K_j \\
\delta_\ve K^j &= - \ol{\n^j} \left( \G^M D_M + \frac32 \tau_0 \Lambda + \frac{1}{4!} \tau_0 \L \psi^{\m\n\r\s} \G_{\m\n\r\s}\right) \Psi
}
Surprisingly enough, we thus conclude that this action is not obtained by summing up the supersymmetric Yang-Mills action \eqref{przaction} with some putative supersymmetric $G_2$-Chern-Simons action, due to the changes in the algebra. In fact, comparing \eqref{t0offshellaction}, \eqref{t0offshellsusy} with \eqref{przaction}, \eqref{przsusy}, we see that other than the addition of the $G_2$-Chern-Simons term, the changes are precisely
\eq{
\tau_0 \rightarrow - \tau_0~,
}
with the exception of the defining Killing spinor equation \eqref{t0ks}. This prohibits any possible interpolation between this action and \eqref{przaction}.

Using \eqref{thesolution}, we reformulate the action \eqref{t0offshellaction} in terms of background fluxes as follows:
\eq{\label{t0backgroundaction}
S  =& \int d^7 x \sqrt{ g_{(7)}}~ \Tr \left( \frac12  F^{MN} F_{MN} - \overline{\Psi} \G^M D_M \Psi - K^j K_j\right) \\
  +& \int d^7 x \sqrt{ g_{(7)}} ~\Tr \left( \frac{1}{56} H^{MNP} \ol{\Psi} \G_{MNP} \Psi
  - \frac{1}{7} H_{MNP} A^M F^{NP}
  - \frac{1}{49} H_{MPQ} H^{NPQ} A_M A^N  \right) \\
  +& i \int   C_3   \wedge \Tr\left( F \wedge F \right)
              + i \int d^7 x \sqrt{ g_{(7)}}~ \Tr  \left( \frac{1}{4!4}  F_4^{MNPQ} \overline{\Psi} \Lambda \G_{MNPQ} \Psi \right)~.
}
A primary motivation for rewriting the action in terms of backgrounds fields rather than torsion is of course the hope to extend the action to the case of more general backgrounds, including more general metrics. However, there are some ambiguities in deciding how to rewrite the action precisely. In particular, note that since $F_4 = 4 i \tau_0 \psi = d C_3$, we have that
\eq{
C_{3\m\n\r} = \frac{1}{4! 4\tau_0} \e_{\m\n\r\s\tau\k\l} F_4^{\s\tau\k\l}~,
}
which is the odd-dimensional selfduality condition for a $(2k-1)$-form in $d= 4k -1$ discussed in \cite{osd}. In addition, the fermionic coupling to the background field $C_3$ can be rewritten as
\eq{\label{osd}
 F_4^{\m\n\r\s} \overline{\Psi} \Lambda \G_{\m\n\r\s} \Psi = - 2 C_3^{\n\r\s} \ol{\Psi} \L \G_{\m\n\r\s} \p^\m \Psi~,
}
thus appearing as a modification of the fermionic kinetic term. We will see that some of these ambiguities are fixed when generalizing the action, as we will now show.

We generalize this procedure to obtain an action on arbitrary co-closed $G_2$-structure manifolds. Given a $Spin(7)$ Killing spinor satisfying $\nabla_\m \chi = \frac12 i \tau_0 \g_\m \chi - \frac18 i \tau_{\m\n} \g^\n \chi$, we see that the $10d$ Killing spinor \eqref{ftks} satisfies
\eq{
\nabla_\m \ve = \frac12 \tau_0 \G_\m \L \ve - \frac18 \tau_{\m\n} \G^\n \Lambda \ve~.
}
As we anticipated, we were unable to extend the pure SYM action \eqref{przaction}, \eqref{przsusy} to spaces with non-trivial $\tau_3$, whereas we have been able to extend the $G_2$-Chern-Simons action \eqref{t0offshellaction}. As deduced in appendix \ref{ftcalc}, we find that the action
\eq{\label{t3offshellaction}
S  = & \int d^7 x \sqrt{ g_{(7)}}~ \Tr \left( \frac12  F^{MN} F_{MN} - \overline{\Psi} \G^M D_M \Psi - K^j K_j \right) \\
& +\tau_0 \int d^{7} x \sqrt{ g_{(7)}} ~ \Tr \left( -  \frac32 \overline{\Psi} \Lambda \Psi  -2 \e_{abc} F^{ab} \phi^c + 8  \tau_0 \phi^a \phi_a \right)\\
            &+ \int  \left( - 4 \tau_0 \psi + \tau_3 \right)  \wedge \Tr\left( A \wedge \d A + \frac23 A \wedge A \wedge A\right) \\
            &+ \int d^7 x \sqrt{ g_{(7)}}~ \Tr
            \left(\frac{1}{4!4} \left(- 4  \tau_0 \psi^{\m\n\r\s} + \tau_3^{\m\n\r\s} \right) \overline{\Psi} \Lambda \G_{\m\n\r\s} \Psi \right)
}
is off-shell supersymmetric with respect to
\eq{\label{t3offshellsusy}
\delta_\ve A_M &= \ol{\Psi} \G_M \ve \\
\delta_\ve \Psi &= \frac12 F_{MN} \G^{MN} \ve - \frac87 \G^{\m\a} \phi_a \nabla_\m \ve - 8 A^\m \left(\nabla_\m + \frac18 \tau_{\m\n}\G^\n \L \right) \ve + \n^j K_j \\
\delta_\ve K^j &= - \ol{\n^j} \left( \G^M D_M + \frac32 \tau_0 \Lambda + \frac{1}{4!} \tau_0 \L \psi^{\m\n\r\s} \G_{\m\n\r\s}\right) \Psi~.
}
In light of \eqref{thesolution}, we see that the action \eqref{t0backgroundaction} does not generalize to the situation with $\tau_3 \neq 0$, due to the sign of $\tau_3$ in \eqref{t3offshellaction}. Indeed, in the presence of $\tau_3$, it follows that the odd selfduality condition \eqref{osd} is no longer valid. Hence, let us rewrite \eqref{t0backgroundaction} as
\al{
S  =& \int d^7 x \sqrt{ g_{(7)}}~ \Tr \left( \frac12  F^{MN} F_{MN} - \overline{\Psi} \G^M D_M \Psi - K^j K_j\right) \\
  +& \int d^7 x \sqrt{ g_{(7)}} ~\Tr \left( \frac{1}{56} H^{MNP} \ol{\Psi} \G_{MNP} \Psi
  - \frac{1}{7} H_{MNP} A^M F^{NP}
  - \frac{1}{49} H_{MPQ} H^{NPQ} A_M A^N  \right) \nn \\
  +  i &\int \fcal_4 \wedge  \Tr\left( A \wedge \d A + \frac23 A \wedge A \wedge A\right)
  +  i \int d^7 x \sqrt{ g_{(7)}} ~\Tr \left(\frac{1}{4! 4} \fcal^{MNPQ} \ol{\Psi} \L \G_{MNPQ} \Psi \right)~. \nn
}
with
\eq{
\fcal_4 = 8 \tau_0 \star_7 \left(C_3 - \frac{1}{8 \tau_0} \star_7 F_4 \right)~,
}
which satisfies $\fcal_4 = F_4$ when $\tau_3 = 0$. We then find that this is the appropriate action which can be generalized to co-closed $G_2$-structure manifolds, in that it matches the action \eqref{t3offshellaction}.

\section{Conclusion}
We have constructed novel $7d$ supersymmetric gauge theories on $G_2$-structure manifolds. Such spaces can be classified according to their torsion classes.
In the case of weak $G_2$-holonomy manifolds, i.e., only non-trivial scalar torsion, we have constructed an alternative to the known
SYM theory \eqref{przaction}, \eqref{przsusy}, namely \eqref{t0offshellaction} \eqref{t0offshellsusy},  which includes a
$G_2$-Chern-Simons term. Even though the $G_2$-Chern-Simons term comes with fermionic counterterms, together these are insufficient
to ensure invariance of supersymmetry and the inclusion of the SYM action appears to be a necessity.  Next,  we have extended this action to co-closed $G_2$-structure spaces, given in \eqref{t3offshellaction}, \eqref{t3offshellsusy}, for which a pure SYM theory is unknown. Examples
of such spaces are $S^3 \times M_4$ with $M_4$ hyperk\"{a}hler and $U(1)$-squashed $S^7$.

These actions were constructed by coupling to a flux background satisfying the supersymmetry constraints of an E7-brane in a supersymmetric type IIA$^*$ background;
obtaining the background by examining an off-shell supergravity background would not have been possible, since the field content corresponds to a non-minimal $7d$ supergravity which is currently unknown off-shell. In addition, whereas previously the additional curvature terms in the action have been somewhat ad hoc,
the nature of the precise coupling terms is reminiscent of a DBI-WZ-like brane theory, thus lending credence to our methods.
The downside is the low amount of supercharges preserved by the theory.  This is independent of the number of admissible Killing
spinors on the underlying manifold, since different spinors lead to different
$G_2$-structures and hence different solutions for $F_4$. Since one of the supercharges has been taken off-shell, the theory is amenable to localization. It would be interesting to compare the results of such computations to the results of pure SYM localization \cite{mz},\cite{prz},\cite{rocen}.

We have used the notion of a kappa-symmetric E7-brane purely as a tool to construct the field theories; similar to the standard procedure of coupling to off-shell supergravity backgrounds, our E7-brane solutions do not solve the equations of motion. However, the violation lies purely in the equation of motion for $F_4$; it may prove worthwhile to turn on fields we have set to zero by hand to see if this can be remedied. Hence from a gravitational point of view, our setup may be viewed as a stepping stone to construct Euclidean brane solutions on curved spaces satisfying the EoM. Another avenue to explore is to look at the second class of brane backgrounds we have obtained, namely the conformally $G_2$-holonomy manifolds with $\tau_{0,2,3} = 0$, $\tau_1 = \d \Delta$. It would be interesting to compare the resulting field theory to recent work on SYM theories on spheres with non-constant couplings \cite{mn}.

\section*{Acknowledgements}
 I would like to thank Achilleas Passias and especially Ruben Minasian for collaboration in various stages of this project. In addition, my thanks to Nikolay Bobev, Fridrik Gautason and Alessandro Tomasiello for useful discussion. I am financially supported in part by INFN and by the ERC Starting Grant 637844-HBQFTNCER.

\appendix
\section{Conventions \& identities}
\subsection{Gamma-matrices}
In this section we give our conventions for the gamma-matrices. We decompose $Spin(1,9) \rightarrow Spin(7) \otimes Spin(1,2)$, with indices
$\m, \n, ... \in \{1,...,7\}$, $a,b,... \in \{0,8,9\}$.
Correspondingly, the 10d gamma matrices we use act on 32-component spinors as\footnote{
For comparison, in our notation the $32 \times 32$ gamma-matrices of \cite{prz} are given by
\all{
\begin{alignedat}{6}
\hat{\g}^{\text{there}}_\m &= \g_\m \otimes - \s_2 \otimes \s_1 ~, &&\qquad &&
\hat{\g}^{\text{there}}_8  &{}={}& \obb  \otimes   \s_1 \otimes \s_1 \\
\hat{\g}^{\text{there}}_9  &= \obb  \otimes   \s_3 \otimes \s_1 ~, &&\qquad &&
\hat{\g}^{\text{there}}_0  &{}={}& \obb  \otimes   \obb \otimes i \s_2 ~.
\end{alignedat}}
As a consequence, $\L^{\text{there}} = \obb \otimes - i \s_2 \otimes \obb$, whereas $\L^{\text{here}} = \obb \otimes \obb \otimes \s_1$.
}.
\eq{\label{gammadecomp}
\G_\m &= \g_\m \otimes \obb \otimes (-\s_2) = i \g_\m \pcal_1\\
\G_a &= \obb \otimes \cg_a \otimes \s_1 = \cg_a \pcal_2\\
\G_{(10)} &= \obb \otimes \obb \otimes \s_3 = \pcal~,
}
where for convenience we have defined
\eq{
\pcal   &= \obb \otimes \obb \otimes \s_3 \\
\pcal_1 &= \obb \otimes \obb \otimes i \s_2 \\
\pcal_2 &= \obb \otimes \obb \otimes \s_1~.
}
As a result, we find the useful relation
\eq{
\g_\m = i  \G_\m \pcal_1 = - i \G_\m \pcal_2 \pcal~.
}
The $3d$ `internal' gamma matrices are defined as $\cg_0 = - i \s_2$, $\cg_8 = \s_3$, $\cg_9 = \s_1$ such that
\eq{
\frac{1}{3!} \e_{abc} \cg^{abc} &= - 1 \\
\frac12 \e_{abc} \cg^{bc} &= - \cg_a \\
\e_{abc} \cg^c &= \cg_{ab} \\
\e_{abc} &= \cg_{abc}~,
}
with $\e_{089} = +1$.
On the brane, which we dub `external space', we use
\eq{
\g_{\m_1 ... \m_7} = - i \e_{\m_1 ... \m_7} ~.
}
The external gamma-matrices are imaginary anti-symmetric.
The charge conjugation matrix decomposes as
\eq{
C_{10} = C_7 \otimes C_3 \otimes \s_1~.
}
Finally, we make use of the definition $\ol{\Psi} = \Psi^T C^{-1}$ for $Spin(1,9)$ spinors.

\subsection{$G_2$-structure}\label{g2}
Our $G_2$-structure conventions match those of \cite{kmt} up to a change of sign for the volume form (and hence, Hodge dual), and match those of \cite{ols1} \cite{ols2} precisely. Let us spell them out explicitly, as well as a number of identities that we require.

The $G_2$-structure is defined by the associate and co-associative forms $( \vf, \psi)$, with $\psi = \star_7 \vf$, which can be expressed in terms of a unimodular Majorana $Spin(7)$ spinor $\chi$ as
\eq{
\vf_{\m\n\r}   &= - i \ol{\chi} \g_{\m\n\r} \chi \\
\psi_{\m\n\r\s} &= -  \ol{\chi}  \g_{\m\n\r\s} \chi~.
}
These satisfy the following useful relations, as can be demonstrated by Fierzing:
\eq{\label{g21}
\psi_{\m\n\k\l} \psi_{\r\s}^{\phantom{\r\s}\k\l} &=  2 \psi_{\m\n\r\s} + 4 \delta_{\m\r} \delta_{\n\s} - 4 \delta_{\m\s} \delta_{\n\r}~,  \\
\psi_{\m\n \k\l} \varphi_{\r}^{\phantom{\rho}\k\l} &=  4 \varphi_{\m\n\r}~, \\
\varphi_{\m\k\l} \varphi^{\n\k\l} &= 6 \delta_\m^\n  \\
\vf_{\k \l \tau} \psi^{\m\n\r \tau} &= - 3 \delta_\k^{[\m} \vf^{\n\r]}_{\phantom{\n\r}\l} + 3 \delta_\l^{[\m} \vf^{\n\r]}_{\phantom{\n\r}\k} ~.
}
In addition, we also have
\eq{\label{g22}
\g_{\m\n}    \chi &= i \varphi_{\m\n\r} \g^\r \chi~, \\
\g_{\m\n\r}   \chi &= i \varphi_{\m\n\r} \chi + \psi_{\m\n\r\s} \g^\s \chi~, \\
\g_{\m\n\r\s}  \chi &= - 4 i \varphi_{[\m\n\r} \g_{\s]} \chi - \psi_{\m\n\r\s} \chi \\
\g_{\m\n\r\s\l} \chi &= -5 \psi_{[\m\n\r\s} \g_{\l]} \chi ~.
}
Existence of a $G_2$-structure ensures that differential forms on $M_7$, which a priori furnish $SO(7)$-representations, can be decomposed into irreducible $G_2$-representations as follows:
\eq{
\O^1 (M_7) &\sim {\bf 1} ~, \qquad
\O^2 (M_7) \sim {\bf 7} \oplus {\bf 14} ~,\qquad
\O^3 (M_7) \sim {\bf 1} \oplus {\bf 7} \oplus {\bf 27} ~.
}
Applying this to the exterior derivatives of the associative and co-associtave forms, we obtain the torsion classes. Specifically, these are defined as
\eq{\label{torsionclasses}
\d \vf  &= 4 \tau_0 \psi + 3 \tau_1 \wedge \vf + \tau_3 \\
\d \psi &= 4 \tau_1 \wedge \psi + \tau_2 \wedge \vf
}
with $\tau_0 \sim {\bf 1}$, $\tau_1 \sim {\bf 7}$, $\tau_2 \sim {\bf 14}$, $\tau_3 \sim {\bf 27}$. Note that there are a number of ways to express the ${\bf 27}$; as a three-form, four-form, or as a traceless symmetric two-form. In \eqref{torsionclasses}, $\tau_3$ is defined to be a four-form. We define
\eq{
\tau_{3\m\n\r\s} = \tau_{[\m}^{\phantom{3[m}\l} \psi_{\n\r\s]r\l}  \quad \iff \quad \tau_{\m\n} = \frac13 \psi^{\r\s\l}_{\phantom{\r\s\l}(\m} \tau_{3n)\r\s\l}~.
}
Using the above, we can find an expression for the connection acting on the spinor in terms of the torsion classes. The basis of spinors is given by $(\chi, \g_\m \chi)$, hence generically one should find $\nabla_\m \chi = A_\m \chi + B_{\m\n} \g^\n \chi$. Decomposing $B \sim {\bf 7} \times {\bf 7} \sim {\bf 1} + {\bf 7} + {\bf 14} + {\bf 27}$ as $B_{\m\n} = a \delta_{\m\n} + \vf_{\m\n\r} b^\r + c_{[\m\n]} + d_{(\m\n)}$, these components can be solved in terms of the torsion classes by constructing various bilinears to obtain the crucial identity
\eq{\label{nablachi}
\nabla_\m \chi = \frac12 i \tau_0 \g_\m \chi - \frac12 i \left( - \vf_{\m\n\r} \tau_1^\r + \frac12 \tau_{2\m\n} + \frac14 \tau_{\m\n} \right) \g^\n \chi~.
}

\section{Closure of the supersymmetry algebra}\label{ftcalc}

\subsection{ $\tau_{1,2,3} =0$}\label{weakclosure}
In this appendix, we will derive the existence of a second realization of the supersymmetry algebra on weak $G_2$-holonomy spaces with corresponding invariant action, in addition to the one discussed in \cite{blau},\cite{fhy},\cite{mz}. Let us consider the following ansatz for the supersymmetry variations:
\eq{\label{t0onshellsusyansatz}
\delta_\ve A_M &= \ol{\Psi} \G_M \ve \\
\delta_\ve \Psi &= \frac12 F_{MN} \G^{MN} \ve + \frac12 (k_3 - 8) \tau_0 \phi^a \L \G_\a \ve - \frac72 k_1 \tau_0 A^M \L \slashed{\psi}\G_M \ve + \frac12 k_2 \tau_0 A^M \L \G_M \ve ~,
}
with $k_{1,2,3} \in \rbb$ parameters to be determined, and $\slashed{\psi} = \frac{1}{4!} \psi^{\m\n\r\s} \G_{\m\n\r\s}$. Making use of the fact that $\slashed{\psi} \ve = - 7 \ve$ due to \eqref{g22}, it can be shown that
\eq{\label{t0susyalg}
\delta_\ve^2 \phi_a &= - v^\n F_{\n a} + [\phi_a, (v^b \phi_b)] + \frac12 (7 k_1 + k_2 + k_3 - 8 ) \tau_0 \left(\ol{\ve} \G_{ab} \L \ve\right) \phi^b  \\
\delta_\ve^2 A_\m   &=- v^\n F_{\n\m} + D_\m (v^b \phi_b) + \frac12 (k_2 - 5 k_1) A^\n \nabla_\m v_\n \\
\delta_\ve^2 \Psi   &= - v^M D_M \Psi - \frac14 \nabla_\m v_\n \G^{\m\n} \Psi + \frac12 (- 4 k_1 - k_3)\tau_0 \left(\Psi \G^a \ve\right) \L \G_a \ve \\
&- \left( \ol{\ve} \left( \G^M D_M - \frac12 (3- k_0 + 4 k_1)  \tau_0 \L + \frac{1}{14} k_0 \tau_0 \L \slashed{\psi} \right) \Psi\right) \ve \\
& + \frac12 v^N \G_N \left(  \G^M D_M - \frac12 (3- k_2 -2  k_1)  \tau_0 \L + \frac12 k_1 \tau_0 \L \slashed{\psi} \right) \Psi \\
& + \frac12 \left(  -2 + \frac12 k_2 - \frac12 k_1 \right) \tau_0 \left(\ol{\ve} \G_{ab} \L \ve\right) \G^{ab} \Psi~,
}
with $v_M = \ol{\ve} \G_M \ve$. Note that when the decomposition of $\ve$ uses only a single $Spin(7)$ Majorana spinor, as will be the case for most of this paper, $v_\m = 0$; we have not made use of this so as to simplify the comparison to the standard SYM algebra. Closure of the algebra enforces
\eq{
k_0 = 7 k_1 ~, \qquad k_2 = 5 k_1 ~, \qquad k_3 = - 4 k_1 ~,
}
such that on-shell $\delta_\ve^2 = - \lcal_v - \gcal_{\phi^a v_a} - R$, with $\lcal_v$ generating a translation along $v$, $\gcal_{\phi^a v_a}$ a gauge transformation with parameter  $\phi^a v_a$ and $R$ an R-symmetry transformation. The fermionic equation of motion is given by
\eq{
\left( \G^M D_M - \frac32 (1-  k_1)  \tau_0 \L + \frac12  k_1 \tau_0 \L \slashed{\psi} \right) \Psi~.
}
Therefore, in order to find an action under this algebra, we take the ansatz
\eq{
S &= \int d^7 x \sqrt{ g_{(7)}}~ \Tr \left( \frac12  F^{MN} F_{MN} - \overline{\Psi} \G^M D_M \Psi \right) \\
+\tau_0 &\int d^{7} x \sqrt{ g_{(7)}}  \Tr \left(   \frac32 (1-k_1) \overline{\Psi} \Lambda \Psi + 2 (1+n_3) F^{ab} \phi^c \e_{abc} + 8 (1+n_2) \tau_0 \phi^a \phi_a \right)\\
            + \tau_0 &\int   \Tr \Big( n_4  \psi  \wedge \left( A \wedge \d A + \frac23 A \wedge A \wedge A\right)
            - \frac{1}{4!2}  k_1\psi^{\m\n\r\s} \overline{\Psi} \Lambda \G_{\m\n\r\s} \Psi \Big)
}
The variation of this action under \eqref{t0onshellsusyansatz} is given by
\al{
\tau_0^{-1} \delta_\ve S ={}&
  \ol{\Psi} \G^{\m\n} \L \ve F_{\m\n} \left(  \frac32 k_1 - n_4 + \frac12 k_1 - 4 k_1 \right)
+ \ol{\Psi} \G^{\m a} \L \ve F_{\m a} \left( 8 k_1 - 3 k_1 - 4 k_1 - k_1 \right) \nn\\
&+ \ol{\Psi} \G^{ab} \L \ve F_{ab}    \left(  8 k_1 + \frac32 k_1 + 6 n_3 - \frac72 k_1 \right)
+ \tau_0 \ol{\Psi} \G^\m \ve A_\m \left(-10 k_1 - 6 (1-k_1) k_1 + 2 k_1^2 \right) \nn\\
&+\tau_0 \ol{\Psi} \G^a \ve \phi_a \left(28 k_1 + 16 n_2 + 12 (k_1^2 - 2 k_1) + 28 k_1 (1-k_1) \right) ~.
}
We thus find two supersymmetrically invariant actions: either $k_1 = n_{2,3,4}  =0$, leading to the known SYM theory \eqref{przaction}, \eqref{przsusy}, or
\eq{\label{kvalues}
k_1 = 2~, \qquad n_2 = 0~, \qquad n_3 = -2 ~, \qquad n_4 = - 4~.
}
This latter choice results in the action
\eq{\label{t0onshellaction}
S &= \int d^7 x \sqrt{ g_{(7)}}~ \Tr \left( \frac12  F^{MN} F_{MN} - \overline{\Psi} \G^M D_M \Psi \right) \\
+\tau_0 &\int d^{7} x \sqrt{ g_{(7)}}  \Tr \left(  - \frac32 \overline{\Psi} \Lambda \Psi - 2 F^{ab} \phi^c \e_{abc} + 8  \tau_0 \phi^a \phi_a \right)\\
            +  &\int   \Tr \Big( -4 \tau_0  \psi  \wedge \left( A \wedge \d A + \frac23 A \wedge A \wedge A\right)
            - \frac{1}{4!}  \tau_0 \psi^{\m\n\r\s} \overline{\Psi} \Lambda \G_{\m\n\r\s} \Psi \Big)
}
invariant under the on-shell algebra
\eq{\label{t0onshellsusy}
\delta_\ve A_M &= \Psi \G_M \ve \\
\delta_\ve \Psi &= \frac12 F_{MN} \G^{MN} \ve - \frac87 \G^{\m\a} \phi_a \nabla_\m \ve - 8 A^\m \nabla_\m \ve ~.\\
}
The algebra can be taken off-shell in the usual way by making use of 7 auxiliary scalar fields $K^j$ and 7 commuting fermions $\nu^j$ satisfying \eqref{nuidentities}.\footnote{We thank Maxim Zabzine for useful comments on the off-shell construction for proper $G_2$-manifolds.}
The construction of such spinors follows along the lines of \cite{evans} and consequently \cite{fhy}.
The action \eqref{t0onshellaction} is invariant under 2 independent supercharges, given by
\eq{
\ve = \chi \otimes \z \otimes \left( \begin{array}{c} 1 \\ 0 \end{array}\right)~,\qquad
\z = e^{\frac34 \pi i} \left( \begin{array}{c} \z_1 \\ \z_2  \end{array}\right)~,
}
where $\chi$ defines the co-associative form $\psi$ in \eqref{t0onshellaction}, $\z_{1,2}$ are real and the phase of $\z$ is fixed by the Majorana condition. Therefore, it follows that $v_\m = 0$ and $v_a v^a = 1$ as remarked in \cite{prz}. In order to take 1 supercharge off-shell, we set $\z_2 = 0$ as a partial R-symmetry fix such that $v_8 = 0$ and $\G_{90} \ve = \ve $. We thus break the bosonic component of the supersymmetry algebra to $g_2 \times so(1,1)$.
By setting
\eq{
\n_j = \G_{j8}\ve~,
}
one finds that \eqref{nuidentities} is satisfied. The off-shell algebra is then given by \eqref{t0offshellsusy}, which leaves the action \eqref{t0offshellaction} invariant. The variation of the auxiliary field squares to
\eq{
\delta_\ve^2 K^j = - v^M D_M K^j - \n^{[j} \left( \G^\m \nabla_\m + \frac32 \tau_0 \L + \slashed{\psi} \right) \n^{k]} K_k ~,
}
where the latter term corresponds to an $SO(7)$ rotation of the scalar fields $K^j$.

\subsection{$\tau_{1,2} = 0$}
We now allow for non-trivial $\tau_{3}$ as well, such that
\eq{
\nabla_\m \ve = \frac12 \tau_0 \G_\m \L \ve - \frac18 \tau_{\m\n} \G^\n \L \ve~.
}
Our ansatz for the supersymmetry algebra will be
\eq{\label{t3onshellsusyansatz}
\delta_\ve A_M &= \ol{\Psi} \G_M \ve \\
\delta_\ve \Psi &= \frac12 F_{MN} \G^{MN} \ve + 4 \tau_0 \phi_a \G^a \L \ve - 4 \tau_0 A^\m \G_\m \L \ve + m A^\m \tau_{\m\n} \G^\n \L \ve
}
with $m \in \rbb$ a parameter to be determined. For $\tau_3 = 0$, the algebra reduces to \eqref{t0onshellsusy}.
Making use of\eqref{t0susyalg} and \eqref{kvalues}, we obtain the following:
\eq{
\delta_\ve^2 \phi_a &= - v^\n F_{\n a} + [\phi_a, (v^b \phi_b)] + 4 \tau_0 \left(\ol{\ve} \G_{ab} \L \ve\right) \phi^b  \\
\delta_\ve^2 A_\m   &=- v^\n F_{\n\m} + D_\m (v^b \phi_b)  \\
\delta_\ve^2 \Psi   &= - v^M D_M \Psi - \frac14 \nabla_\m v_\n \G^{\m\n} \Psi + \tau_0 \left(\ol{\ve} \G_{ab} \L \ve\right) \G^{ab} \Psi \\
&- \left( \ol{\ve} \left( \G^M D_M + \frac32 \tau_0 \L + \tau_0 \L \slashed{\psi} - \frac14 \L \slashed{\tau_3}\right) \Psi\right) \ve \\
& + \frac12 v^N \G_N \left(  \G^M D_M + \frac32 \tau_0 \L + \tau_0 \L \slashed{\psi} - \frac14 \L \slashed{\tau_3}\right) \Psi \\
& + m \left(  - \frac12 v^N \G_N \L  \slashed{\tau_3} \Psi  -\frac14 \left(\ol{\ve} \G_\n \G^\r \L \ve\right) \tau_{\m\r} \G^{\m\n} \Psi\right)
}
Therefore, the unique closed extension of the algebra to spaces with non-trivial $\tau_3$ is given by $m = 0$.

Next, we look for an action invariant under the algebra \eqref{t3onshellsusyansatz}. Taking into account that the fermionic equation of motion should correspond to
\eq{
\left(  \G^M D_M + \frac32 \tau_0 \L + \tau_0 \L \slashed{\psi} - \frac14 \L \slashed{\tau_3}\right) \Psi~,
}
our ansatz will be to extend \eqref{t0onshellaction} as
\eq{
S &= \int d^7 x \sqrt{ g_{(7)}}~ \Tr \left( \frac12  F^{MN} F_{MN} - \overline{\Psi} \G^M D_M \Psi \right) \\
& +\tau_0 \int d^{7} x \sqrt{ g_{(7)}}  \Tr \left(   -\frac32 \overline{\Psi} \Lambda \Psi  - 2 \e_{abc} F^{ab} \phi^c + 8  \tau_0 \phi^a \phi_a \right)\\
            & + \tau_0 \int   \Tr \Big(- 4  \psi  \wedge \left( A \wedge \d A + \frac23 A \wedge A \wedge A\right)
            - \frac{1}{4!}  \psi^{\m\n\r\s} \overline{\Psi} \Lambda \G_{\m\n\r\s} \Psi \Big) \\
            & + \int \Tr \left(  m_1 \tau_3 \wedge \left( A \wedge \d A + \frac23 A \wedge A \wedge A\right)
            + \frac{1}{4!4} m_2 \tau_3^{\m\n\r\s} \Psi \L \G_{\m\n\r\s} \Psi \right)~,
}
with $m$, $m_1$, $m_2$ to be determined by supersymmetry invariance. Using
\eq{
\lcal_{CS2}^F &= - \frac{1}{4!} \overline{\Psi}\L \G_{\m\n\r\s} \Psi \ti{\tau_3}^{\m\n\r\s}\\
\implies \delta_\ve \lcal_{CS2}^F &= \ol{\Psi} \G_\s \ve \vf^{\m\n\r} \tau_{\m}^{\phantom{\m} \s} F_{\n\r}
+ 2 \ol{\Psi} \G^{\m a} \L \ve \tau_\m^{\phantom{\m}\n} F_{\n a}
- 8 \tau_0 \ol{\Psi} \G^\m \ve \tau_{\m\n} A^\n
}
we obtain
\al{
\delta S ={}&
  \ol{\Psi} \G^{\m\n} \L \ve    \tau_{\m}^{\phantom{\m}\r} F_{\r \n} \left(\frac12 - \frac12 m_1 + \frac12 m \right)
+ \ol{\Psi} \G^{\m a} \L \ve    \tau_{\m}^{\phantom{\m}\r} F_{\r a}  \left(\frac12 - \frac12 m_2             \right)\nn\\
&+\ol{\Psi} \G_\s \ve \vf^{\m\n\r} \tau_\m^{\phantom{\m}\s} F_{\n\r} \left(\frac14 m_1  - \frac14 m_2 \right)
+ \tau_0 \ol{\Psi} \G^\m \ve \tau_{\m\n} A^\n                        \left(-2 + (- 3 - 2 + 5) m + 2 m_2\right)\nn \\
&+\ol{\Psi} \G_\m \ve  \tau^{\m\n} \tau_{\n\r} A^\r \left(\frac12 - \frac12 m_2  \right) m ~.
}
Thus, the unique solution is given by
\eq{
m_1 = m_2 =1 ~, \qquad m = 0 ~.
}
The fact that $m\stackrel{!}{=}0$ for invariance of the action matches the solution to the constraint coming from closure of the algebra.

As explained in appendix \ref{weakclosure}, one of the two supercharges can be taken off-shell, leading to the algebra \eqref{t3offshellsusy} which preserves the action \eqref{t3offshellaction}. The closure of the algebra when acting on the auxiliary scalar $K^j$ is satisfied with
\eq{
\delta_\ve^2 K^j &=  - v^M D_M K^j - \n^{[j} \left(\G^\m \nabla_\m  + \frac32 \tau_0 \L + \L \slashed{\psi} - \frac14 \slashed{\tau_3} \right) \n^{k]} K_k~.
}

\end{document}